\documentclass[aps,prl,preprint,showpacs]{revtex4}

\usepackage{graphicx}
\usepackage{dcolumn}

\begin{document}

\title{Quantum Conductance in Silver Nanowires:\\correlation between atomic structure and transport properties}

\author{V. Rodrigues$^{1,2}$}
\author{J. Bettini$^{1}$}
\author{A.R. Rocha$^{1,2}$}
\author{L.G.C. Rego$^{1}$}
\author{D. Ugarte$^{1}$}
\email{ugarte@lnls.br}
\affiliation{$^{1}$Laborat\'{o}rio Nacional de Luz S\'{\i}ncrotron, C.P. 6192, 13084-971 Campinas SP, Brazil}
\affiliation{$^{2}$Instituto de F\'{\i}sica "Gleb Wataghin", UNICAMP, C.P. 6165, 13083-970 Campinas SP, Brazil}

\date{\today}

\begin{abstract}
We have analyzed the atomic arrangements and quantum conductance of silver nanowires generated by 
mechanical elongation. The surface properties of Ag induce unexpected structural properties,
as for example, predominance of high aspect ratio rod-like 
wires. The structural behavior was used to understand the Ag quantum conductance data and 
the proposed correlation was confirmed by means of theoretical calculations. These results emphasize that the conductance of 
metal point contacts is determined by the preferred atomic structures and, that atomistic descriptions are essential 
to interpret the quantum transport behavior of metal nanostructures. 
\end{abstract}

\pacs{68.65.+g, 61.16.Bg, 73.50.-h}

\maketitle

The physical interpretation of the presence of flat plateaus and abrupt jumps on the electrical conductance 
of metal nanowires (NWs) \cite{RuintenbeekBook} has been controversial, because its discrete nature has been either 
attributed to electron channels \cite{Olesen,Krans} or to structural rearrangements \cite{KransPRL}. Since the beginning 
of this research field, the basic experiment for studying the NW conductance has been based on the elongation 
of junctions or point contacts \cite{LandmanScience}, while measuring their electrical properties. In fact, it has been 
extremely difficult to discriminate between structural and electronic effects in a NW elongation experiment, 
because both are simultaneously modified during the measurement \cite{Rubio}.

Recently, the application of time-resolved High Resolution Transmission Electron Microscopy (HRTEM) in 
the study of gold NWs has revealed the existence of suspended atom chains (ATCs), whose conductance 
was measured to be equal to the universal quantum, $G_{0} = 2 e^{2}/h$  (where $e$ is the electron charge and $h$ is Planck's constant) \cite{Ohnishi}. Subsequently, a broader correlation between atomic 
structure and conductance has been derived for Au NWs \cite{PRLAu} where a simple extension of the Wulff construction \cite{Marks} was used to 
predict the NW morphologies based on the crystallographic directions. For example, it is well-known that the Au 
compact (111) facets are preferred and, then, nanosystems evolve to expose mainly these low energy facets
\cite{Marks}.

Although significant progress has been done to understand NW properties, most of the reported
studies are actually based on gold (see recent review in \cite{RuintenbeekBook}). Moreover, the majority of the available
evidence is derived indirectly on the basis of average statistical behaviors \cite{Krans,PRLAu}. 
To get a deeper and more general insight it is necessary to analyze still other model systems, as for example, 
different monovalent metals (to easily describe the conductance) and, if possible, with different surface properties (in order 
to induce different NW structures). Silver represents an excellent point in case, because it is a face centered cubic (fcc) metal with a lattice 
parameter similar to that of gold, nonetheless, the silver minimal energy facets are (100) oriented \cite{RefAgfacets}. 
It must be emphasized, that the higher reactivity of silver has hindered detailed studies of quantum conductance (QC) in such 
system \cite{RefAg_QC}. 

In this work, we have used two independent experimental methods to analyze silver NW structures and their 
conductance, finding that the observed Ag NW properties differ strongly 
from the previously studied systems. Theoretical predictions have been used to consistently correlate the structural 
and QC behavior. The results 
provide new information to understand the conductance properties of nanosystems. 

We have generated the silver NWs \textit{in situ} in a HRTEM using the procedure reported by Kondo and Takayanagi 
\cite{Kondo}, who were able to produce NWs by making holes in a metal self-supported thin film. The silver film 
was polycrystalline (5 nm thick, average grain size 50-100 nm); a detailed 
description of this experimental procedure was given previously \cite{ PRLAu, RodriguesEuro}. The HRTEM 
observations were performed using a JEM 3010 URP with a 0.17 nm point resolution (LME/Lab. Nac. Luz S\'{\i}ncrotron, 
Campinas, Brazil). All imaging work was acquired close to Scherzer defocus \cite{Ref15}, and the presented images 
were generated by the digitalization of video recordings (30 frames/s) acquired using a high sensitivity 
TV Camera (Gatan 622SC). This kind of \textit{in situ} HRTEM NW study has been previously performed with high efficiency 
on gold and platinum \cite{Ohnishi,PRLAu,Kondo,RodriguesEuro,Takai,TakayanagiHelical,PRBATC,PRLPt}. 
However, silver NWs have been much more 
difficult to image with similar quality. Firstly, due to the lower atomic number the generated contrast is 
much weaker; secondly, Ag NWs display a much faster evolution, rendering difficult the 
real time image acquisition.

The electrical conductance of silver NWs was measured with an independent and dedicated instrument: 
a Mechanically Controllable Break Junction (MCBJ) 
\cite{Muller} operating in ultra-high-vacuum (UHV) \cite{UHVMCBJ} ($<$ 10$^{-8}$ Pa). In this method, 
a silver wire ($\phi$  = 75 $\mu m$, 99.99 $\%$ pure) is broken 
\textit{in situ} in UHV and NWs are generated by putting into contact and subsequently retracting 
these clean surfaces. The electronics is basically composed of a home-made voltage source and a current-voltage 
converter coupled to an 8 bits digital oscilloscope (Tektronic TDS540C). The acquisition system input 
impedance and time response were optimized to perform conductance measurements in the [0-4] $G_{0}$ range with a relative 
error of $\Delta$\textit{G/G} $\sim$ 10$^{-4}$. 

The time-resolved HRTEM observations of silver NWs have revealed some striking and unexpected structural behavior, which 
contrast strongly with the results reported for gold \cite{Kondo,Ohnishi,PRLAu,PRBATC, RodriguesEuro}. In particular, 
high aspect-ratio rod-like NWs 
along the [110] direction (hereafter noted as [110] NW) are the most frequently observed morphology. As an 
example, in Fig. 1, we show a series of snapshots of a complete elongation/thinning process of a pillar shaped 
NW. Initially, the rod is formed by 5 (200) atomic planes (thickness $\sim$ 0.8 nm, Fig. 1a), losing sequentially one 
atomic plane at a time to attain a three layer thickness ($\sim$ 0.4 nm, Fig. 1c). Subsequently in Fig. 1d, the right side of the wire 
becomes thicker, while the left side maintains the 
same width ($\sim$ 0.4 nm) as in Fig. 1c. However, this left sector shows a quite different contrast pattern, the HRTEM image shows 
darker dots for the external planes (tube-like), whereas the central layer contrast is much weaker. Finally, before 
breaking, the minimal observed size 
for this morphology consists of two (200) atomic planes ($\sim$ 0.2 nm wide, Fig. 1e). All the lattice fringes and angular relations observed in the HRTEM images of Ag NWs can be fully described by means of the bulk silver fcc structure. It is important to note that NWs showing the tube-like 
contrast pattern are frequently observed in the experiments and the underlying atomic structure seems to be 
particularly stable because they can attain aspect-ratios $>$ 10 (see example in Fig. 1g). In addition, the final length 
of these tube-like NWs is not determined by the initial size, as evidenced in 
Fig. 1a-f. In fact, in many cases, we have observed that when the wire attains this peculiar structure (or contrast 
pattern), the apexes retraction cause the NW to elongate by a factor $\sim$ 1.5-3 without thinning. This lengthening  
reflects the enhanced strength of this atomic configuration. 

\begin{figure}
\includegraphics[width = 8.5 cm]{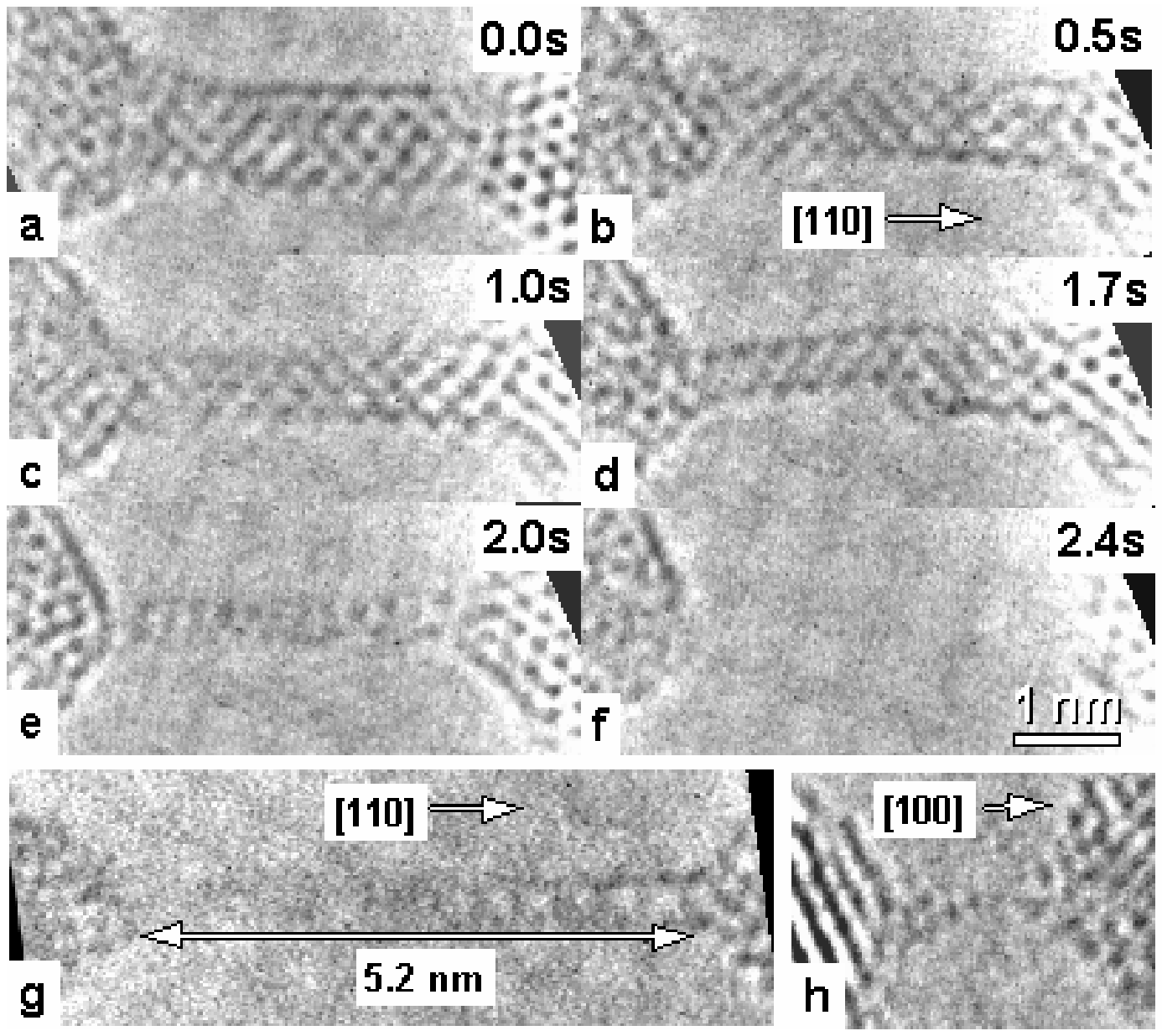}
\caption{a-f) Atomic resolution micrographs showing the elongation and thinning of a rod-like silver nanowire (see text for explanations); g) 
HRTEM image of a high aspect-ratio pillar like NW (width $\sim$ 0.4 nm and length is $\sim$ 5 nm); h) HRTEM micrograph of silver suspended atom chain. Atomic positions appear dark.}
\end{figure} 

As for the existence of silver suspended atom chains, our results confirm that they do occur, nevertheless 
they are much less frequently observed than in our previous studies of gold and platinum NWs 
\cite{PRBATC,PRLAu,PRLPt,RodriguesEuro}. In fact, they are only seen when 
one of the junction apexes is oriented along a [100] direction. These ATCs are 2-4 atoms long with a bond length in the 0.33-0.36 nm range. Both the lengths and 
bond distances are similar to previous reports on Au and Pt ATCs 
\cite{Ohnishi,Yanson,PRBATC,PRLAu,PRLPt}. However an important 
difference has been revealed by the dynamic HRTEM recordings, which have pointed out that silver [111] NWs 
display a fast and abrupt rupture preventing the formation of ATCs (within our time resolution). 

In order to deduce the three dimensional atomic arrangement of the nanowire from the HRTEM images (basically a 
bidimensional projection), the geometrical Wulff construction \cite{Marks} can be used. This approach yields the crystal 
shape by predicting the relative size of the lower energy facets of the crystal. Recently, it has been successfully applied 
to describe and model gold nanojunctions \cite{PRLAu}. 
Fig. 2 shows the application of the Wulff method to model the morphology of silver rod-like NWs. It is instructive 
to look first at the expected morphology of a silver nanoparticle, a truncated cuboctahedron with regular triangular (111) facets, 
where the relevance of (100) facets can be easily identified (see Fig. 2a)\cite{Blair}. The cross section of a [110] silver NW 
can be derived by looking at this cuboctahedron along the [110] axis. Figs. 2b and 2c show the suggested cross-sections for 
the rod-like NWs seen in Figs. 1a and 1c, which are formed by 5 and 3 (200) atomic planes, respectively. These 
rods are generated by the alternate stacking of two different planes containing 11 and 8 atoms (marked 11/8, and 
displayed with different colors in Fig. 2b) for the thicker NW and 4/3 atoms for the thinner one (Fig. 2c). When these 
rods are observed along a [1-10] axis (as in the experiment), we observe the bidimensional projection indicated as side 
views in Fig. 2. In a first approximation, at the Scherzer defocus \cite{Ref15} and for such a thin object, the expected 
contrast at each atomic column position should be proportional to the projected atomic potential or, in other words, 
the number of atoms along the observation direction. In Fig. 1d, the NW contrast is tube-like, with the external 
planes much darker than the central one. Thus, on the light of the preceding argument, the central atomic columns 
should contain less atoms than the border ones. The only way to fulfill these conditions, by thinning the 4/3 rod 
(Fig. 2c), is to build a 4/1 NW (see Fig. 2d). Finally, the thinnest Ag [110] 
NW, shown in Fig. 1e, consists of two atomic planes. Two atomic arrangements may yield the observed contrast: a 2/1 structure 
with a triangular cross-section 
(Fig. 2e), or two parallel atom chains marked as 1/1 in Fig. 2f. The signal-to-noise ratio in the image does not 
allow us to identify which one is observed in Fig. 1e. Recently, Hong \textit{et al} \cite{AgNWScience} have reported the 
preparation of Ag [110] rod-like NWs within porous matrix
by means of wet chemical methods. However the generated structure (which can be described as 2/2 following our notation) 
has not been observed in the free standing Ag NWs studied here.

\begin{figure}
\includegraphics[width = 8.5 cm]{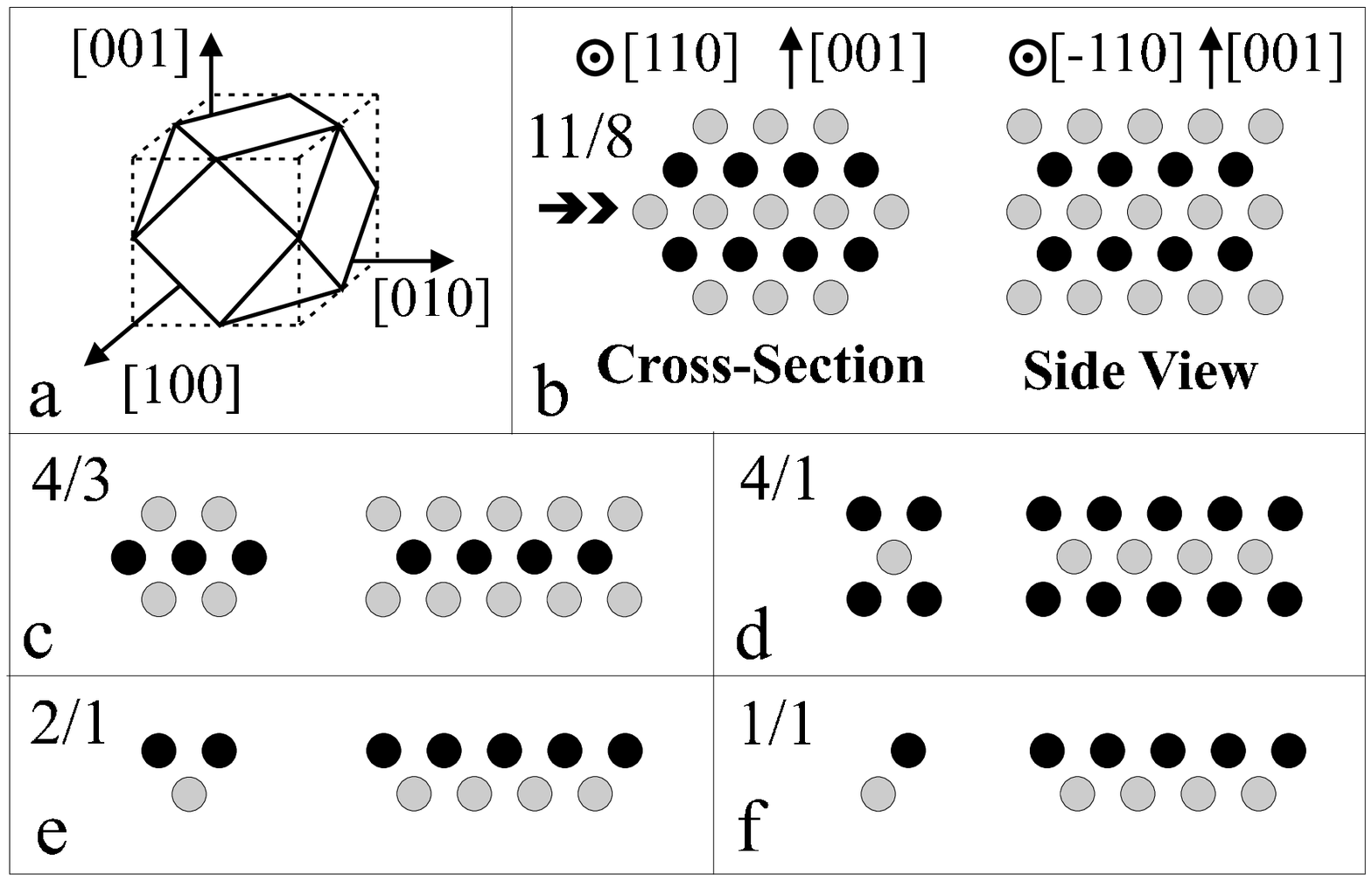}
\caption{a) Application of the Wulff construction \cite{Marks} to determine the shape of silver nanoparticles. 
b-f) Scheme of possible atomic arrangements for rod-like Ag NWs; we show the NW cross-sections and also, 
side views (observed along the horizontal arrow in b).}
\end{figure} 

The discussion presented above allows the deduction of the atomic structures for the NW morphology that have been 
most frequently observed during the HRTEM studies. It is now tempting to correlate the preferred atomic arrangements 
with the quantum conductance properties. Fig. 3 shows typical conductance curves of Ag NWs (see inset) and, 
a histogram of occurrence of each conductance value (global histogram \cite{RuintenbeekBook})
obtained from 500 conductance curves. This histogram shows large peaks at 1 $G_0$, $\sim$ 2.4 $G_0$ and $\sim$ 4 $G_0$. It also displays major 
differences when compared to similar results reported for gold NWs \cite{RuintenbeekBook}. 
In the first place, the 1 $G_0$
peak is not the dominant one \cite{RuintenbeekBook,PRLAu}, in accordance with the HRTEM data that shows low occurrence of atom chains. 
Secondly, the absence of the 2 $G_0$ peak (usually located at $\sim$ 1.8 $G_0$ in Au histograms \cite{RuintenbeekBook}) evidences that 
the structure of silver NWs should be much different. In addition, the large peak at 2.4 $G_0$ displays an area 
approximately 2.5 times bigger than the peak close to 1 $G_0$. This fact suggests that this conductance is associated with a more frequently 
occurring structure, which has to be identified as the rod-like [110] NWs observed in our HRTEM experiments. Because the minimal 
observed [110] Ag NW should consist of two atomic layers (Fig. 2e), it is tempting to associate the conductance 
peak at $\sim$ 2.4 $G_0$ with this structure.

\begin{figure}
\includegraphics[width = 7 cm]{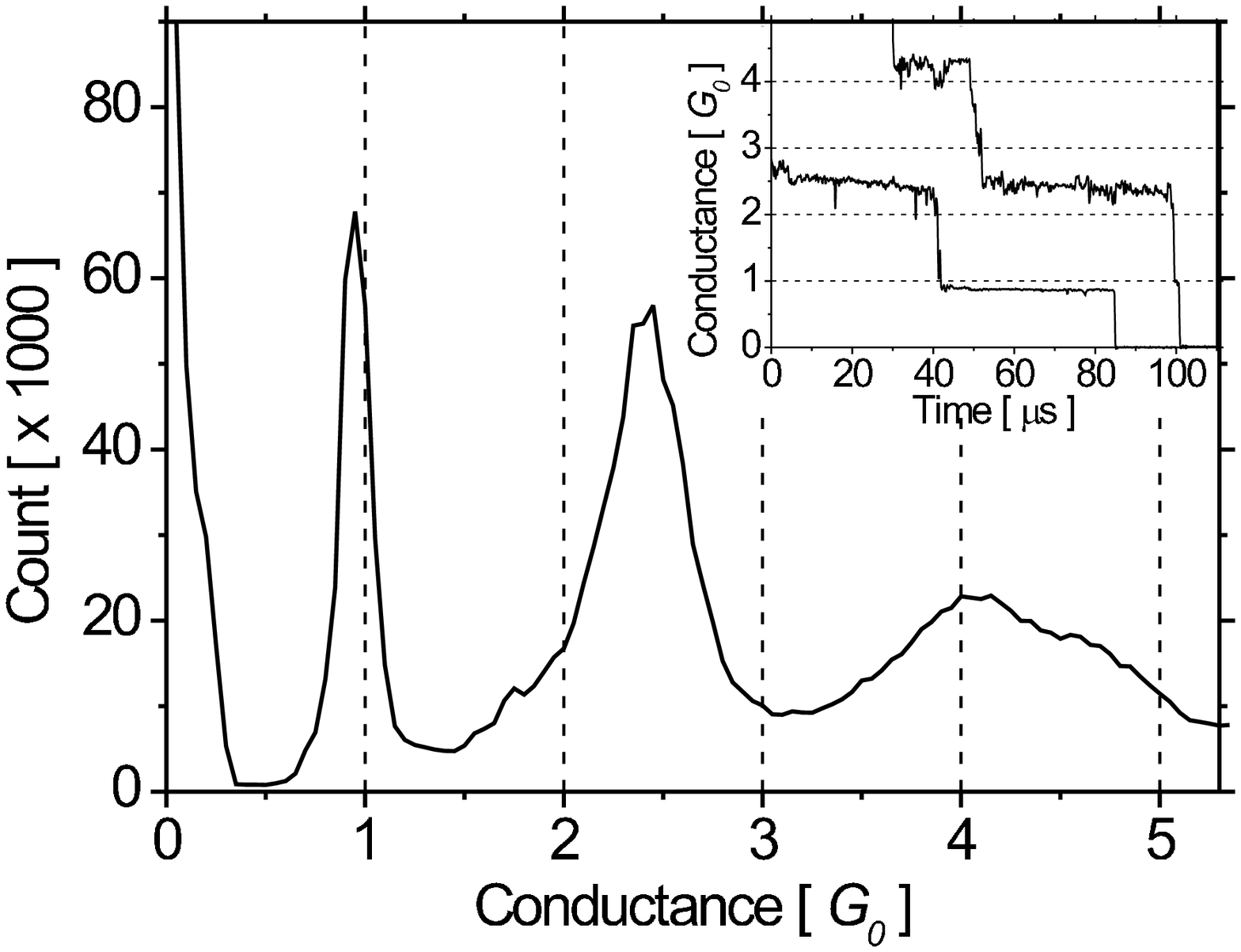}
\caption{Global histogram of conductance from Ag NWs.
The main features are the peaks located at 1 $G_0$, 2.4 $G_0$ and 4 $G_0$. Inset: Typical conductance curves;
note the plateaus connected by abrupt jumps.}
\end{figure}

Finally, to consistently correlate the experimental results of the conductance and structural behaviors, we have performed theoretical 
conductance calculations for the structures presented in Fig. 2. Based on the experimental data, it is clear that 
a proper theoretical description should take into account the 
atomic arrangement of the NW. For this purpose we have used an approach introduced by Emberly and 
Kirczenow \cite{Emberly} that is based on the Extended-H\"{u}ckel-Theory, the latter being employed to obtain the molecular 
orbitals (MO) of the Ag NWs \cite{EHT}. The MO calculations take into account the s, p and d orbitals of the Ag atoms in 
the NW, as well as overlap and energy 
matrix elements extending beyond the first neighbor atoms. The electronic transport was 
described within the Landauer scattering formalism \cite{RuintenbeekBook}. A whole description of 
the procedure has been presented elsewhere \cite{LRego}. Fig. 4 shows theoretical calculations of the conductance for 
the different NW morphologies observed in the HRTEM images. The experimental results are to be compared 
with the conductance at the Fermi energy ($E_F$), which is indicated by the vertical line in the figure.
The conductance curve oscillations in Fig. 4 are sensitive to the atomic positions, therefore an average 
of the conductance around $E_F$ yields a more representative value of $G$ that takes 
into account the atomic vibrations during the measurement.

\begin{figure}
\includegraphics[angle = -90,width = 8.5 cm]{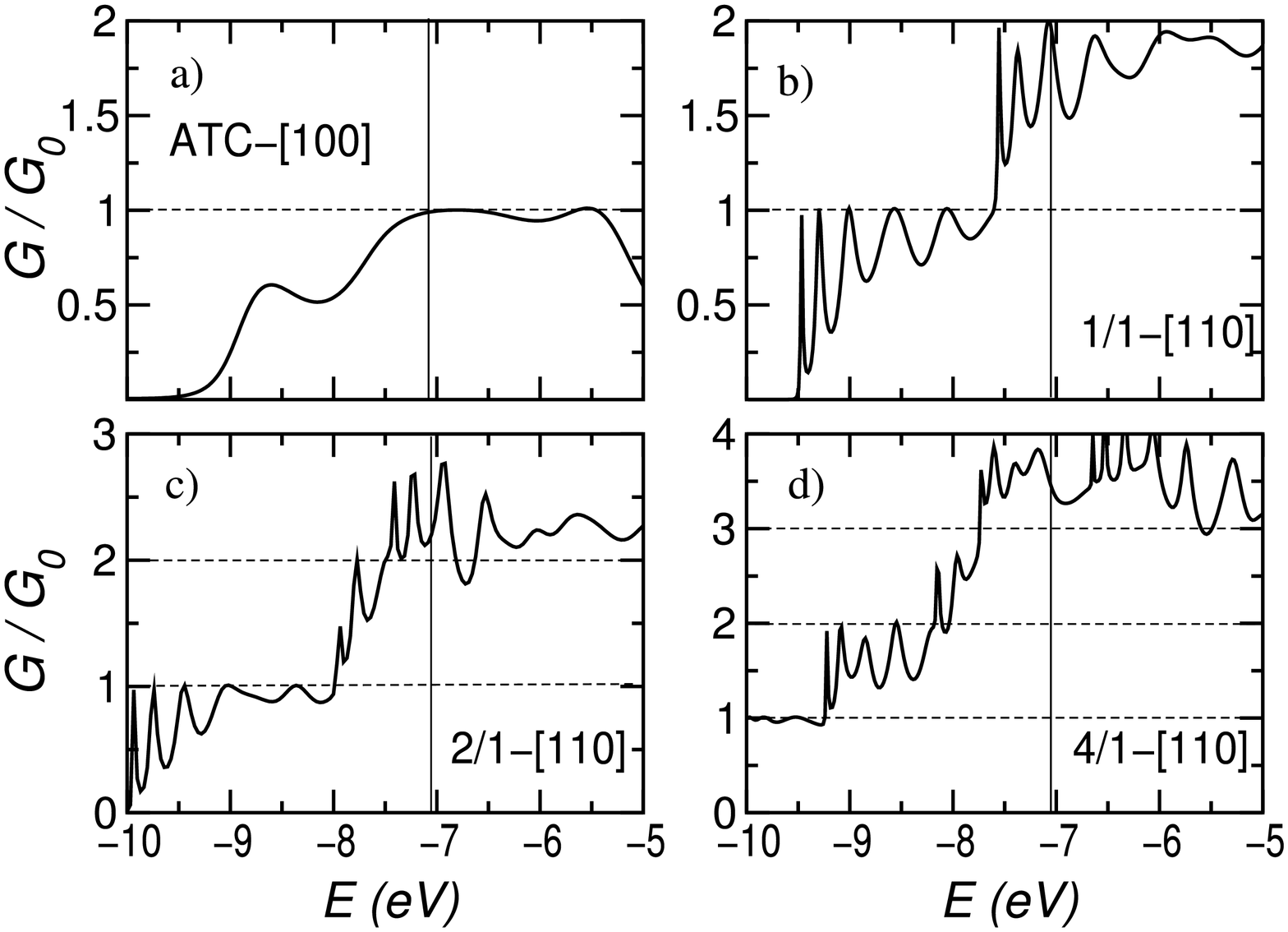}
\caption{ Theoretical calculations of the quantum conductance $G$ as a function of the electron energy for different Ag NW morphologies:
2 atoms long ATC in the [100] direction; 1/1 (b), 2/1 (c) and 4/1 (d) rod-like structures along the [110] direction. The conductance 
is plotted in units of $G_0$ and the vertical line indicates the Fermi energy.}
\end{figure}  

The calculations show that [100] Ag ATCs (interatomic distance 0.29 nm) display the expected conductance close to 
the quantum $G_0$ (Fig. 4a) \cite{Ohnishi,PRLAu}. The 1/1 rod, composed of two parallel atom chains, shows a conductance close 
to 1.8 $G_0$ (Fig. 4b), however, the Ag global histogram (Fig. 3) does not show a peak associated with this value 
and it must be concluded that 
this atomic arrangement does not occur in our experiments. As for the 2/1 [110] NW the model predicts a conductance 
at $G(E_F)$ $\sim$ 2.5 $G_0$ that is in remarkable agreement with the main peak of the conductance histogram. 
These results show that the 2/1 
structure is in fact the minimal rod-like silver NW (Fig. 2e). Finally, the conductance calculation for the 4/1 rod 
with rectangular cross section yields $G(E_F)$ $\sim$ 3.8 $G_0$, which should be associated with the observed conductance peak at 
$\sim$ 4 $G_0$. All the theoretical conductance values show an excellent agreement with the experiments, confirming 
the proposed correlation between HRTEM images and conductance histogram. 

In summary, we have been able to reveal the preferred structures of silver NWs generated by mechanical elongation and determine 
the conductance for each kind of NW. This correlation between structural and electronic properties was confirmed by 
means of theoretical calculations. These results represent a clear evidence of the need to determine precisely the 
atomic arrangement of NWs in order to analyze in detail their conductance behavior. Although the surface properties 
of silver suggested that Ag NWs should be quite different from gold ones, the 4/1 or 2/1 rod-like wire 
would have been rather difficult to predict. This fact emphasizes the importance of experiments allowing the direct 
determination of atomic arrangements in the field of nanosystems. 

The authors are grateful to FAPESP, CNPq, and LNLS for financial support


\begin{thebibliography}{}

\bibitem{RuintenbeekBook}
J.M. van Ruitenbeek in \textit{Metal Clusters at Surfaces}, edited by K.-H. Meiwes-Broer, Cluster Series (Springer-Verlag Berlin Heidelberg, New York, 2000).
\bibitem{Olesen}
L. Olesen, \textit{et al}., Phys. Rev. Lett. \textbf{72}, 2251 (1994).
\bibitem{Krans}
J.M. Krans \textit{et al}., Nature \textbf{375}, 767 (1995).
\bibitem{KransPRL}
J.M. Krans \textit{et al}., Phys. Rev. Lett. \textbf{74}, 2146 (1995).
\bibitem{LandmanScience} 
U. Landman \textit{et al.}, Science \textbf{248,} 454 (1990).
\bibitem{Rubio} 
G. Rubio, N Agrait and S. Vieira, Phys. Rev. Lett. \textbf{76}, 2302 (1996).
\bibitem{Ohnishi}
H. Ohnishi, Y. Kondo, and K. Takayanagi, Nature \textbf{395}, 780 (1998).
\bibitem{PRLAu}
V. Rodrigues, T. Fuhrer and D.Ugarte, Phys. Rev. Lett. \textbf{85}, 4124 (2000).
\bibitem{Marks}
L.D. Marks, Rep. Progr. Phys. \textbf{57}, 603 (1994).
\bibitem{RefAgfacets}
S.M. Foiles, M.I. Baskes, M.S. Daw, Phys. Rev. B \textbf{33}, 7983 (1986).
\bibitem{RefAg_QC}
J.L. Costa-Kr\"{a}mer \textit{et al.} in \textit{Nanowire}, edited by P.A. Serena and N. Garc\'{\i}a,
NATO ASI Series E: Apllied Science, Vol. \textbf{340} (Kluwer Academic Publishers, The Netherlands, 1997).
\bibitem{Kondo}
Y. Kondo and K. Takayanagi, Phys. Rev. Lett. \textbf{79}, 3455 (1997).
\bibitem{RodriguesEuro}
V. Rodrigues and D. Ugarte, Europ. J. Phys. D \textbf{16}, 395 (2001).
\bibitem{Ref15}
D.B. Williams and C.B. Carter, \textit{Transmission Electron Microscopy} (Plenum Press, New York, 1996) p. 465.
\bibitem{Takai}
Y. Takai \textit{et al.}, Phys. Rev. Let. \textbf{87}, 106105 (2001).
\bibitem{TakayanagiHelical}
Y. Kondo and K. Takayanagi, Science \textbf{289}, 606 (2000).
\bibitem{PRBATC}
V. Rodrigues and D.Ugarte, Phys. Rev. B. {\bf 63,} 073405 (2001).
\bibitem{PRLPt}
V. Rodrigues and D. Ugarte, submitted.
\bibitem{Muller} 
C.J. Muller, J.M. van Ruitenbeek and L.J. de Jongh, Phys. Rev. Lett. {\bf 69,} 140 (1992).
\bibitem{UHVMCBJ}
V. Rodrigues, Master thesis, Universidade Estadual de Campinas, 1999; V. Rodrigues and D. Ugarte, Rev. Sci. 
Instrum., submitted.
\bibitem{Yanson}
A.I. Yanson \textit{et al.}, Nature \textbf{395}, 783 (1998).
\bibitem{Blair}
B.D. Hall, M. Fl\"{u}eli, R. Monot and J.-P. Borel, Phys. Rev. B. {\bf 43,} 3906 (1991).
\bibitem{AgNWScience} 
B.H. Hong \textit{et al.}, Science \textbf{294,} 348 (2001).
\bibitem{Emberly} 
E.G. Emberly and G. Kirczenow, Phys. Rev. B {\bf 58,} 10911 (1998); ibidem {\bf 60,} 6028 (1999).
\bibitem{EHT} 
J.P. Lowe, {\it Quantum Chemistry} (Academic,NY,1978). For a numerical implementation of the method:
G.A. Landrum and W.V. Glassy, YAeHMOP project, http://yaehmop.sourceforge.net.
\bibitem{LRego}
L.G.C. Rego \textit{et al.}, in preparation.

\end{thebibliography}
\end{document}